\documentclass[twocolumn,prl,showpacs,multicol,amsmath,amssymb]{revtex4-1}
\usepackage{amssymb}
\usepackage{amsmath}

\usepackage{graphicx}
\usepackage{epstopdf}
\usepackage{bm}

\graphicspath{{Figures/}}

\newcommand{\boldeta}{{\bm \eta}}

\newcommand{\be}{\begin{equation}}
\newcommand{\ee}{\end{equation}}

\newcommand{\bea}{\begin{eqnarray}}
\newcommand{\eea}{\end{eqnarray}}
\newcommand{\bd}{\begin{displaymath}}
\newcommand{\ed}{\end{displaymath}}
\newcommand{\ba}{\begin{array}}
\newcommand{\ea}{\end{array}}

\def\bk{{\bf k}}

\def\bq{{\bf q}}

\def\bd{{\bf d}}
\def\bz{{\bf z}}

\def\UPT{\rm{UPt$_3$}}
\def\CCI{\rm{CeCoIn$_5$}}

\begin{document}
\date{\today}
\title{Surface state tunneling signatures in two-component superconductor \UPT}

\author{Fabian Lambert$^1$}
\author{Alireza Akbari.$^{2,3}$}
\author{Peter Thalmeier$^4$}
\author{Ilya Eremin$^1$}

\affiliation{$^1$Institut f\"ur Theoretische Physik III, Ruhr-Universit\"at Bochum, 44801 Bochum, Germany}
\affiliation{$^2$Asia Pacific Center for Theoretical Physics, Pohang, Gyeongbuk 790-784, Korea}
\affiliation{$^3$Department of Physics, and Max Planck POSTECH Center for Complex Phase Materials, POSTECH, Pohang 790-784, Korea}
\affiliation{$^4$Max Planck Institute for Chemical Physics of Solids, D-01187 Dresden, Germany}

\begin{abstract}
Quasiparticle interference (QPI) imaging of Bogoliubov excitations in quasi-two dimensional unconventional superconductors has become a powerful technique for measuring the superconducting gap and its symmetry.  Here, we present the extension of this method to three-dimensional superconductors and analyze the expected QPI spectrum for  the two-component heavy fermion superconductor \UPT~whose  gap structure is still controversial. Starting from a 3D electronic structure and the three proposed chiral gap models $E_{1g,u}$ or $E_{2u}$, we perform a slab calculation that determines the 2D continuum Bogoliubov- de Gennes (BdG) surface quasiparticle bands and in addition the in-gap flat-band Andreev bound states that lead to surface Weyl arcs connecting the projected gap nodes. Both features are very distinct for the three models, in particular the most prominent $E_{2u}$ candidate is singled out by the existence of {\it two} Weyl arcs due to the double monopole node points. The signature of these distinct surface bound and continuum states that is left in QPI is derived and discussed. We show that it provides a fingerprint that may finally determine the true nodal structure of \UPT~superconductor.
\end{abstract}


\maketitle

The Cooper-pairing mechanism in many heavy-fermion superconductors (SCs) has not yet been identified. Partially this is connected to the fact that the superconducting gap symmetry, which encodes this mechanism, was not yet unambiguously determined in these compounds. Among various experimental techniques, Bogoliubov quasiparticle interference (QPI) as measured by the scanning tunneling spectroscopy imaging has 
become a notable technique for studying 
gaps of SCs, which possess strong quasi-two-dimensional electronic structure such as high-T$_c$ cuprates \cite{mcelroy:03} and Fe-pnictides \cite{hanaguri:10}. 

However, if the quasiparticle energy has a significant dispersion along $k_z$ direction, the resulting  Fermi surface (FS) also shows considerable corrugation along $k_z$ as it is for example the case in \CCI. Then one could either use an effective artifical 2D FS model \cite{allan:13,vandyke:14} or a model with corrugation and then integrate over the momentum perpendicular to the plane \cite{akbari:11}. Both are ad-hoc methods that cannot be applied to fully 3D SCs  like 
\UPT.

The heavy fermion metal \UPT~is the only known unconventional SC \cite{fisher:89} with a two-component gap function. It must therefore belong to an $E$-type representation of its hexagonal $D_{6h}$ symmetry group.
The early thermodynamic evidence  like two specific heat jumps \cite{fisher:89,trappmann:91,vorenkamp:93} and two upper critical field curves \cite{bruls:90,adenwalla:90,vandijk:93} is reviewed in Refs.~\cite{sauls:93,joynt:02,thalmeier:05}. Finally a consensus emerged that the $E_{2u}$ triplet $f$-wave gap function \cite{norman:96,graf:00} rather than the originally proposed \cite{hess:89,joynt:90} singlet $E_{1g}$ model is realized. Both model gap functions have similar nodal structure: A line node in the hexagonal $ab$ symmetry plane and two point nodes at the poles on the $k_z$-axis which are of first ($E_{1g}$) or second ($E_{2u}$) order. More recently  as a result of thermal conductivity \cite{machida:12,tsutsumi:12,izawa:14} and specific heat \cite{kittaka:13} measurements in rotating magnetic field a revision to a triplet $E_{1u}$ model was controversially discussed. It has again first order point nodes at the poles but node lines are located in two {\it off-symmetry} planes parallel to hexagonal $ab$ plane. In the low temperature, low field B-phase of \UPT~both order parameter components are finite with an intrinsic phase shift $\pi/2$. Therefore  this chiral $B$-phase exhibits a time-reversal symmetry (TRS) breaking \cite{strand:09,schemm:14} condensate with angular momentum $L_z=1$ ($E_{1g,u}$) or  $L_z=2$ ($E_{2u}$).

In this letter, we present the extension of quasiparticle interference method to the 
three-dimensional SC \UPT. In particular, we show that QPI remains a very powerful tool for three-dimensional electronic bands and allows to resolve the bulk nodal gap structure of \UPT. 
We perform appropriate finite slab calculation for the electronic structure to  compute the QPI spectrum of \UPT~at the various surface terminations. Furthermore, 
we produce the correct surface state structure, which is due to the non-trivial momentum space topology of the three proposed gap functions as discussed recently in Refs. \onlinecite{tsutsumi:13,kobayashi:14,goswami:15,kobayashi:16}. The node line is a vortex in {\bf k}-space which will lead to zero energy (flat band) Andreev bound states that correspond to Majorana fermions \cite{sato:16}. The first order node points are single ($E_{1g,u}$) \bk-space monopoles with Chern numbers ${\it C}=\pm L_z =\pm 1$ or double  ($E_{2u}$) \bk-space monopoles with Chern numbers ${\it C}=\pm L_z =\pm 2$. The bulk quasiparticle excitations at the node points are massless isotropic ($L_z=\pm1$) or more general anisotropic ($L_z=\pm2$) Weyl fermions. They lead to zero energy Andreev bound states or Weyl arcs along the surface projection of a path connecting the (anti-) monopole node points \cite{goswami:15}.  Our slab calculations are able to derive explicitly the flat bands due to bulk line nodes and Weyl arcs due to point nodes. Finally we show that the momentum resolved QPI spectrum should exhibit the signature of projected bulk quasiparticle bands for finite bias voltage as well as the topological surface states at zero bias. Both should be distinct due to the different bulk nodal structures of gap candidates. Therefore QPI may provide a direct fingerprint to finally distinguish between candidate $E$-type gap functions that have been discussed for so long, in a similar way that it was able to resolve the ambiguity of $d$-wave singlet gap functions   that persisted before in \CCI~\cite{akbari:11,allan:13,zhou:13}.

%
\begin{figure}
\centering
\includegraphics[width=1\linewidth]{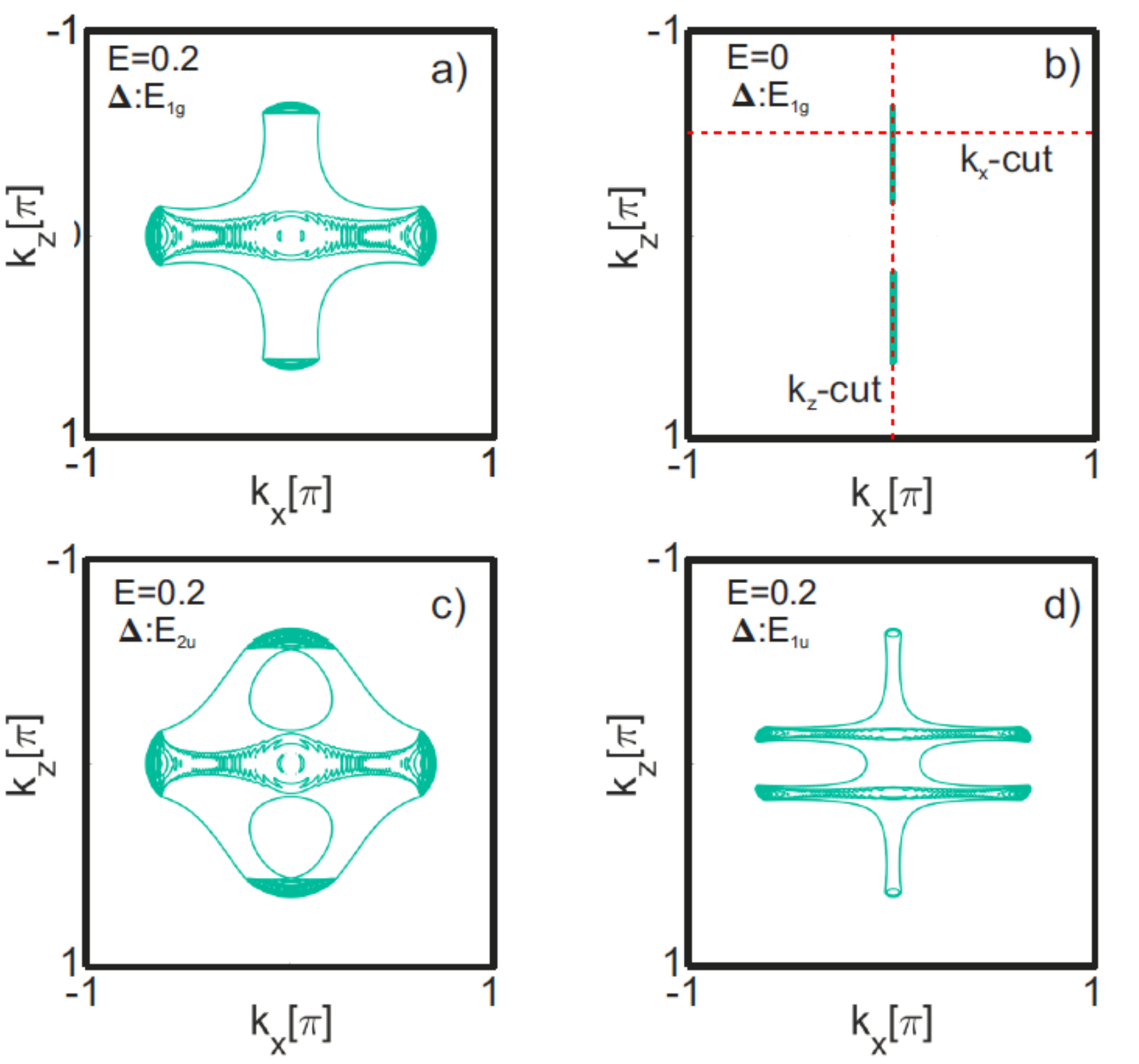}
\vspace{-0.91cm}
\caption{
(Color online)
Equal energy contours $E_\nu(\bk_\parallel)=E$  in $k_xk_z$-plane  for  $E_{1g}$ (a) $E_{2u}$ (c) and $E_{1u}$  (d), with $E=0.2 (\Delta_0=0.05 \epsilon_0)$. Top and bottom pockets are due to polar point nodes, central (a,c) or off-center (d) bands due to line nodes of SC gap. (b) shows zero energy surface bound state (twofold degenerate Majorana state) for $E=0$ ($E_{1g}$).
FS parameters (in meV) are: $t_\parallel = 1.45, t_\perp =1.6, \epsilon_0=9, \mu = 4.5$.}
\label{fig:Fig1}
\end{figure}
%
\begin{figure}
	\begin{center}
\includegraphics[width=0.99\columnwidth]{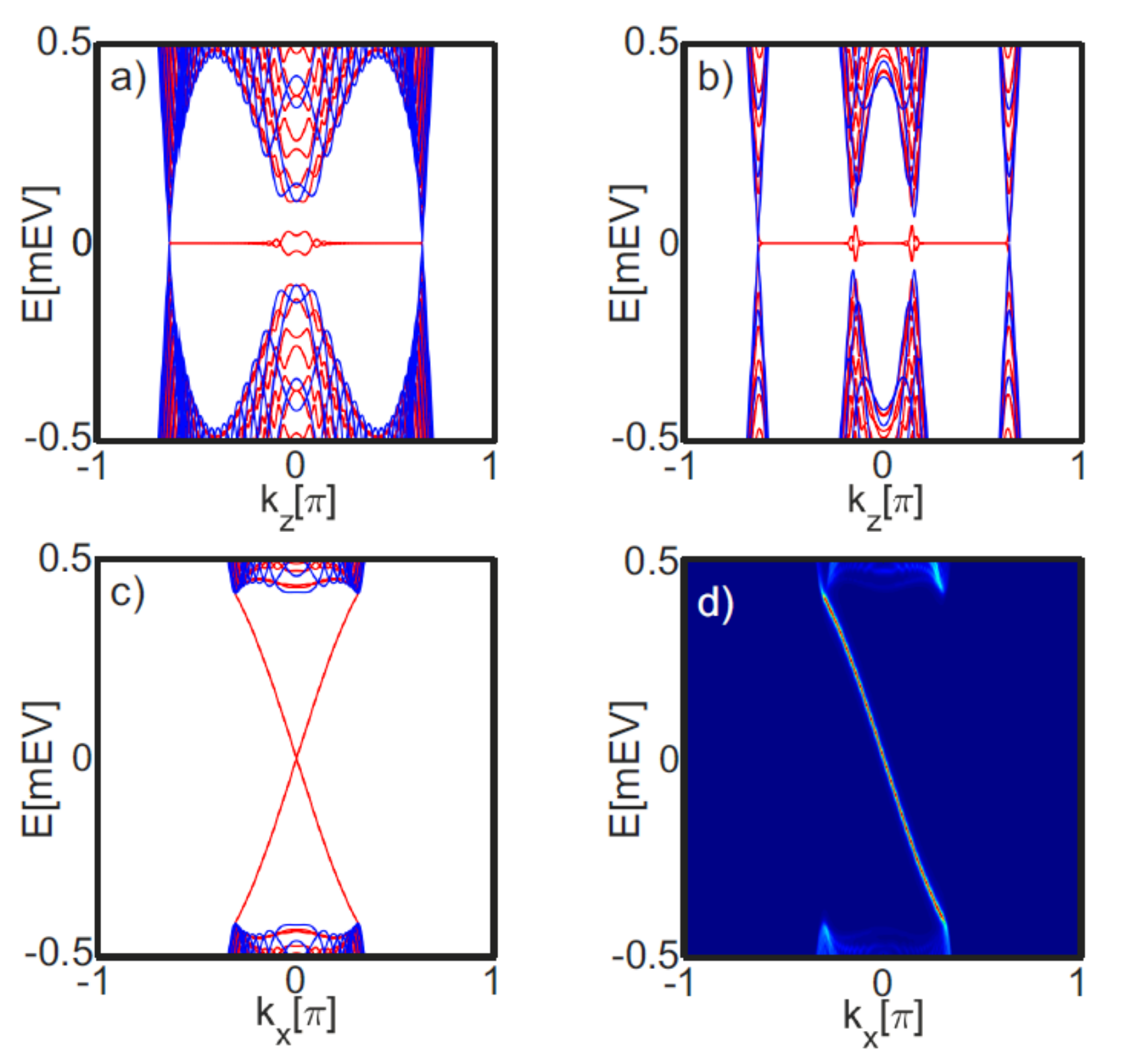}
\end{center}
\vspace{-0.81cm}
\caption{
(Color online)
Band structures for in $k_x$- and  $k_z$ cuts as indicated in Fig.~\ref{fig:Fig1}(b) for $E_{1g}$(a,c) and $E_{1u}$(b). In the continuum region the $k_xk_y$ plane projected bulk bands (blue) agree well with the slab calculation (red). The latter also delivers the topological in-gap surface states: Dispersionless 2-fold degenerate Majorana-type flat bands along $k_x$ (a,b) and dispersive surface states along $k_z$ crossing the gap (c). In (d) the surface density of gap states in (c) is shown for the top surface. At the bottom surface the mirror image $(k_x\rightarrow -k_x)$ is obtained.
}
\label{fig:Fig2}
\end{figure}
%
\begin{figure}
	\begin{center}
%
\includegraphics[width=1\columnwidth]{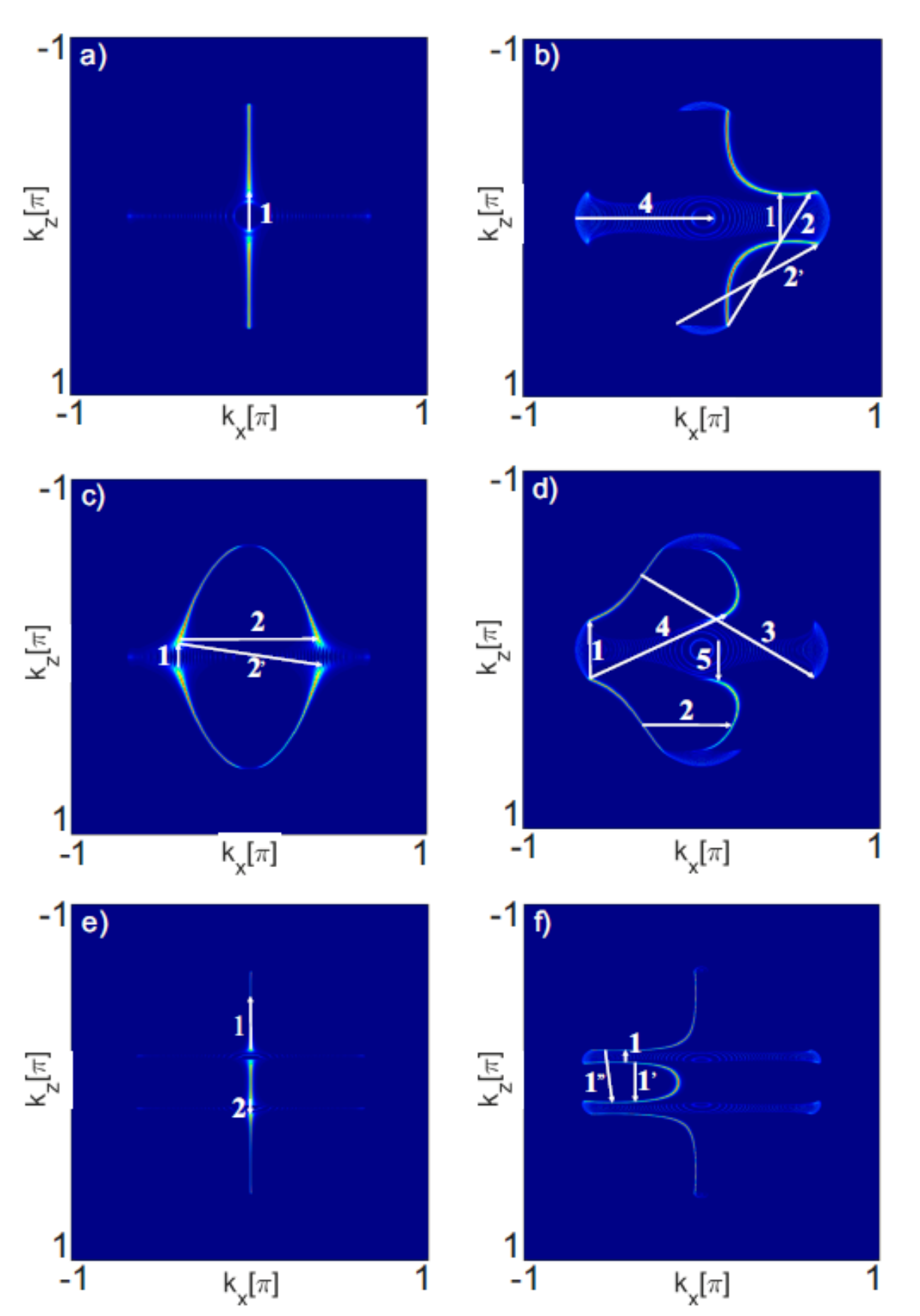}
\end{center}
\vspace{-0.7cm}
\caption{
(Color online)
Left: surface DOS $N_S^0(\bk_\parallel E=0)$ in $k_xk_z$ contour plot with Weyl arcs 
for $E_{1g}(a)$, $E_{2u}(c)$ and $E_{1u}(e)$.
 In (c) two arcs appear due to double monopole point nodes at $(0,0,\pm k_z^0)$. Arc interruption occurs when crossing a node line once in (a),(c) or twice in (e). Right: surface DOS 
contours for $E=0.2$ at top layer. The DOS is reflected $(k_x\rightarrow -k_x)$ for $E\rightarrow -E$.
Characteristic wave vectors $\bq_i, \bq'_i \cdots$ featuring prominent in QPI spectrum (Fig.\ref{fig:Fig4}) are denoted by $1,1' \cdots$.}
\label{fig:Fig3}
\end{figure}
%
%

{\it Model and Methods:}
The 3D heavy bands of \UPT~with a multisheet 
FS \cite{taillefer:88,zwicknagl:02,mcmullan:08,gannon:15,nomoto:16} may be approximated by a global ellipsoid FS \cite{wu:02,gannon:15} with an average in-plane/c-axis effective mass ratio $m_\parallel/m_\perp\simeq 3.4$ \cite{gannon:15}. Note, however that the masses of the main heavy $\Gamma$- centered $\Gamma_3$ sheet which may dominate QPI are $m_\parallel = 80m_e$ and $m_\perp= 101m_e$ \cite{kimura:98,mcmullan:08} with the opposite anisotropy $m_\parallel/m_\perp=0.79$.
For numerical calculation it is preferable to use the approximate ellipsoid from a periodic band model (we set lattice constants $a,c=1$) 
$\xi_\bk=\epsilon_0-2t_\parallel(\cos k_x +\cos k_y) +2t_\perp\cos k_z -\mu$ with $t_\parallel=\hbar^2/2m_\parallel$, $t_\perp=\hbar^2/2m_\perp$ and $\epsilon_0=2(2t_\parallel +t_\perp)$ and $\mu$ denoting the chemical potential. For convenience of presentation $k_{x,y}, k_z$ will be rescaled such that the FS is spherical; i.e. the effective mass anisotropy is eliminated.
The SC candidate models of \UPT~are described by singlet $(\psi_i)$ and triplet  $(\bd_i)$ gap functions $\Delta_\bk=\eta_1\psi_1(\bk)+\eta_2\psi_2(\bk)$ or $\bd_\bk=\eta_1\bd_1(\bk)+\eta_2\bd_2(\bk)$.  Here $\psi_i(\bk), \bd_i(\bk);\; (i=1,2)$ are the even or odd basis functions of the twofold orbitally degenerate singlet or triplet state, respectively. In the low temperature chiral B-phase both orbital components of the in-plane complex gap vector $\boldeta = (\eta_1,\eta_2)$ are non-vanishing and are given by  $\boldeta = \Delta_0(1,\pm i)$. In the case of the odd parity triplet candidates, the Knight shift results \cite{kitaoka:00} indicate that at zero field the \bd-vector of triplet pairs might be weakly pinned along the $c$-direction due to a small pseudo spin-orbit coupling, i.e. $\bd(\bk)=d_z(\bk)\hat{\bz}\equiv \Delta_\bk\hat{\bz}$. An additional component along $b$-axis due to the background AF small moment order \cite{tou:98} is neglected here. With appropriate basis functions the gap models are then given by 
\begin{eqnarray}
\begin{aligned}
\text{E}_{1g}:\Delta_\bk&={\Delta_0}k_z(k_x \pm ik_y),
\\
\text{E}_{2u}:\Delta_\bk&={\Delta_0}k_z(k_x \pm i k_y)^2,
\\
\text{E}_{1u}:\Delta_\bk&={\Delta_0} (5k^2_z-1)(k_x \pm ik_y).
\label{gapFunctionsTrans}
\end{aligned}
\end{eqnarray}
Here we only consider the chiral $E_{1u}$ state that breaks  TRS  \cite{izawa:14}.
For the numerical calculations it will be preferable to use a periodic form with $k_i\rightarrow \sin k_i$.
In the triplet functions the pseudo-spin corresponds to the Kramers degeneracy of the quasiparticle bands enforced by inversion and TRS. 
Under such condition the approximate pseudo-spin rotational symmetry allows for triplet gap functions with stable line nodes \cite{kobayashi:14,kobayashi:16}.

 Now we briefly sketch the calculation of quasiparticle surface density of states (DOS) and STM-QPI spectrum in a finite slab geometry. For a 3D FS this poses a fundamental problem. The tunneling process itself and the impurity scattering of quasiparticles leading to the detectable local DOS ripples on the surface have a purely 2D character. On the other hand while scattering from surface impurities the electrons may also propagate into the bulk and back again due to the perpendicular hopping element $t_\perp$. This leads to a sampling or averaging of 3D dispersion effects in the purely 2D Fourier transformed QPI spectrum where only momenta parallel to the surface occur. Previously such 
3D effects were not treated adequately. For this purpose one has to introduce a finite slab model. We consider slabs assembled from layers $n= 1~\cdots ~N$, parallel to $(k_xk_z)$-plane which contain both the node line and node points. Then only momenta $\bk_\parallel$ within $(k_xk_z)$-plane are defined and the Hamiltoninan is constructed in terms of $c_\bk^{\dagger} =\sum_y c_{\bk_{\parallel}}^{\dagger}\left(y\right) e^{-ik_yy}$ electron operators with $y=n/N$ the layer coordinate and $\bk=(\bk_\parallel,k_y)$. In terms of the layer Nambu spinors 
$\Psi_{\bk_{\parallel}}^{\dagger}\left(n\right)=
\Bigl(c_{\bk_{\parallel},\uparrow}^{\dagger}\left(n\right),
c_{\bk_{\parallel},\downarrow}^{\dagger}\left(n\right),
c_{-\bk_{\parallel},\uparrow}\left(n\right),
c_{-\bk_{\parallel},\downarrow}\left(n\right) \Bigr)$
the Hamiltonian of the superconducting slab is written as
\begin{eqnarray}
\begin{aligned}
&
H=\sum_{k_{\parallel},n,m} \Psi_{k_{\parallel}}^{\dagger}\left(n\right) H_{Sk_{\parallel}}^{n,m} \Psi_{k_{\parallel}}\left(m\right),
\\
&
H_{Sk_{\parallel}}^{n,m}= \sum_{k_y}e^{
ik_y
\frac{(n-m)}{N}
}
\begin{pmatrix}
\xi_\bk \otimes \sigma_0 & \hat{\Delta}_\bk \\
\hat{\Delta}_\bk^{\dagger} & -\xi_\bk \otimes \sigma_0
\end{pmatrix}
,
\end{aligned}
\end{eqnarray}
with the singlet or triplet gap matrix $\hat{\Delta}_\bk=\Delta_\bk(i\sigma_y)$ or  $\hat{\Delta}_\bk=\Delta_\bk\sigma_x$, respectively. 
The eigenvalues $E_\nu(\bk_\parallel)$ $(\nu=1 \cdots \mbox{N})$ of the slab Hamiltonian matrix $\hat{H}_{S\bk_\parallel}=\{H^{n,m}_{S\bk_\parallel}\}$ give the projected surface bulk states and the possible surface bound states,
 which decay exponentially into the bulk (along $y$).
The DOS of quasiparticles is then obtained from the slab Green's function: 
$\hat{G}_S^0(\bk_\parallel,E)=(E+i\delta-\hat{H}_{S\bk_\parallel})^{-1}$ as $N_n^0(\bk_\parallel,E)=(-1/\pi){\rm Im}Tr  [P_\tau\hat{G}_{Snn}^0(\bk_\parallel,\omega)]$ for the $n$-th layer. Explicitly we have for the surface (top, $n=1$) layer, using the Nambu projector $P_\tau$:
\begin{equation}
N_S^0(\bk_\parallel,E)=-\frac{1}{\pi}
{\rm Im}
 Tr
 \Big[
 \frac{1}{2}(\tau_3+\tau_0)\otimes\sigma_0G^0_{S11}(\bk_\parallel,E)
 \Big].
\end{equation}
The impurity scattering potential causing QPI is restricted to few surface layers and assumed diagonal in the layer index, i.e. $V^{nm}_S=V_n\delta_{nm}$ with $V_n=V_0\lambda_n$ where $\lambda_n$ describes the layer dependence and $V_0$ is the momentum-independent scattering strength. Then the  scattering T-matrix is  
$\hat{T} (\omega) = \hat{V}_S\left(1 - \hat{V}_S \sum_{\bk_{\parallel}}\hat{G}_S^0\left(\bk_{\parallel},\omega \right)\right)^{-1}$.
This leads to the 2D slab QPI spectrum corresponding to the local density oscillation of $N_0 \ll N$ surface layers (bias voltage $\omega=eV$ suppressed on r.h.s.):
\begin{eqnarray} 
\begin{aligned}
\delta N_S(\bq_\parallel,\omega)
&=-\frac{1}{\pi}
{\rm Im}
\sum_{\bk_\parallel}\sum_{n=1}^{N_0}\sum_{m=1}^{N} 
\\
&
\hspace{-1.2cm}
\Bigl[G_S^{0n,m}(\bk_\parallel)
T_mG_S^{0m,n}(\bk_\parallel+\bq_\parallel)  \Bigr]
- \Bigr[ \bq_\parallel \rightarrow -\bq_\parallel \Bigr]
,
\end{aligned}
\end{eqnarray}
where $N_0$ is the depth of experimental sampling, while $N$ is in principle the whole sample thickness but is effectively limited by the decay of the surface impurity scattering $(\lambda_n)$ into the bulk.\\

%
\begin{figure}
	\begin{center}
\includegraphics[width=1\columnwidth]{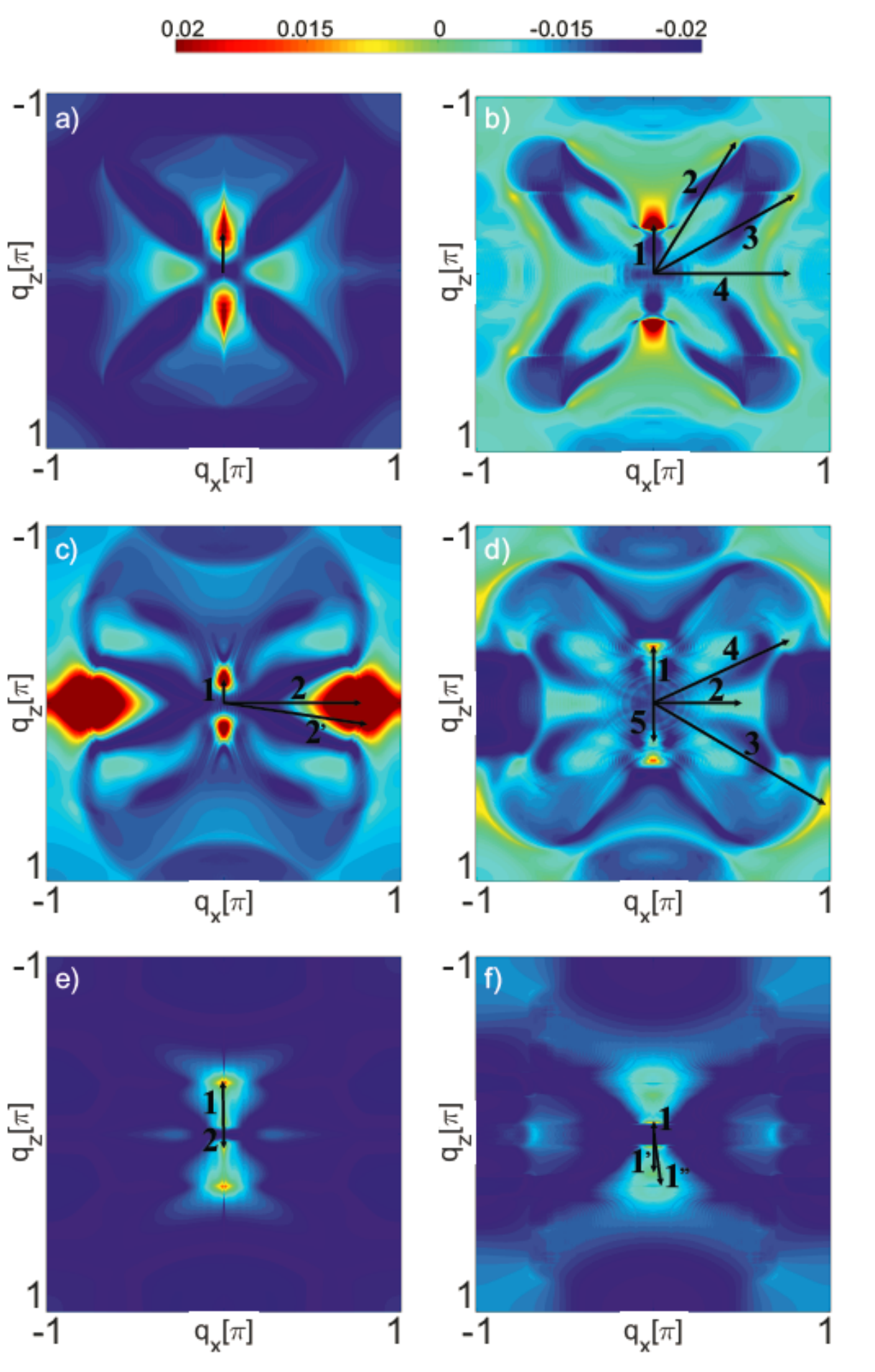}
\end{center}
\vspace{-0.5cm}
\caption{
 (Color online)
 Comparison of QPI spectrum for $E_{1g}$ (top), $E_{2u}$ (center) and $E_{1u}$ (bottom) at zero bias  ($\omega=0$, left column) and positive bias  ($\omega=0.2$, right column). Left column gives the QPI image of topological  Weyl arcs in Fig.
\ref{fig:Fig3}, right column corresponds to continuum surface states QPI contribution. Vectors $\bq_i,\bq_i' \cdots $ of characteristic QPI structures are denoted by $1,1' \cdots$ (c.f. Fig~\ref{fig:Fig3}). Here we use $N_0=6$ and a Heaviside function $\lambda_n=\Theta\left(N_0-n\right)$ as the impurity potential decay.}
\label{fig:Fig4}
\end{figure}
%

{\it Numerical results:} The equal energy contours for eigenvalues of the slab BdG Hamiltonian are shown in Fig.~\ref{fig:Fig1}(a,c,d) for the three gap functions for nonzero energy. Their nodal structures can be clearly discerned. For $E_{1g}$(a) and E$_{2u}$(c)  the polar nodes lead to top and bottom pockets and the central band shaped feature is due to the equatorial node line. Their relative sizes are different due to first or second order point nodes, meaning that for $E_{2u}$ the point node pockets are comparatively large. Most prominently $E_{1u}$ has {\it two} nodal bands around non-symmetry positions along $k_z$, furthermore pockets due to point nodes are small due to large gap velocity in $k_z$ direction. These distinct quasiparticle spectral features of the three gap candidates are also expected to be seen in the QPI images.

In addition the slab calculations give the surface bound states and flat bands enforced by the non-zero topological charges of Weyl node points and node line. In the $E_{1g}$ case this leads to the appearance of a 2D FS (Fig.\ref{fig:Fig1}b) i.e. the Weyl arc connecting the projections of polar topological nodes at $(0,0,\pm k_z^0)$ to the $k_xk_z$ plane where $k_z^0= \cos^{-1}[(-2t_\parallel+t_\perp)/2t_\perp]$. 

The surface state character can be seen in the slab dispersion in two orthogonal $k_z$ and $k_x$- cuts as shown in Fig.~\ref{fig:Fig2}. In the $k_z$ cut (a) the continuum region from the slab calculation (red) and the projected 3D bulk band (blue) are identical. Inside the SC gap a new suface flat band appears in the slab calculation which creates the 2D Weyl arc FS. 
At the nodal line crossing ($k_z=0$ for $E_{1g}$) the zero energy flat band shows the small hybridization gap. 
For $E_{1u}$ (b) there are two such crossings.
In the $k_x$ (c) cut the surface arc states disperse rapidly away form the nodal connection line and traverse the gap, finally merging into the bulk states. The corresponding surface DOS $N_S^0(\bk_\parallel,E)$ of the top surface is shown in a contour plot (d). Note that only one (the left-moving) surface state appears in the DOS, the right moving is located on the bottom surface.
The zero energy flat bands which correspond to BdG twofold degenerate  Majorana fermion states are present in all  three gap models leading to the Weyl arcs in $(k_xk_z)$ plane. Depending on the topological charge there are one ($E_{1g},E_{1u}$) or two ($E_{2u}$) Weyl arcs to be expected. This is confirmed by our explicit calculations which show the shape and density of zero enery modes $(E=0)$ along the arcs (Fig.~\ref{fig:Fig3}, left column). When an arc crosses a node line the zero energy modes are gapped out due to hybridization with nodal bulk states. Therefore each Weyl arc breaks up into two ($E_{1g}$, $E_{2u}$) (a,c), or three ($E_{1u}$) (e) segments depending on the number of bulk node lines.

For sufficiently large energies $(E=0.2)$ the surface DOS is dominated by continuum states (c.f. Fig.~\ref{fig:Fig2}). This is shown in Fig.~\ref{fig:Fig3} (right column) for the three gap functions and the top surface layer in the $k_xk_z$ plane. For $E < 0$ or for the bottom layer the DOS is reflected with respect to $k_x$ coordinate. The DOS contours follow qualitatively the shape of the equal energy contours in Fig.~\ref{fig:Fig1}. However the surface DOS value is suppressed precisely in the projected nodal regions with a large number of bulk states of same energy  due to the hybridization effect. The surface DOS plots from slab calcualtion may be used as a foundation for understanding the QPI images in a similar way as usually done for purely 2D systems: One may try to identify special characteristic scattering wave vectors $\bq^i_\parallel$ that connect points of high DOS in Fig.~\ref{fig:Fig3}  and will therefore appear prominently in QPI image $\delta N_S(\bq_\parallel,\omega)$. A few possible dominant scattering vectors are indicated in Fig.~\ref{fig:Fig3} (right column). It is obvious that for zero (Weyl arcs) as well as for finite (continuum states) bias voltage the surface DOS is profoundly distinct for the three chiral gap models.

Therefore it is to be expected that the QPI spectrum gives a correspondingly different fingerprint for the three cases as shown in  Fig.~\ref{fig:Fig4} for zero bias (left column) and $\omega =0.2$ (right column). The former is the QPI image of Weyl arcs and displays the striking difference between gap functions with different number of arcs leading to large QPI intensity around $\bq=(\pm\pi,0)$ for $E_{2u}$ (c) due to inter-arc scattering which is absent for  $E_{1g,u}$ (a,e). This observation alone would, e.g., be able to distinguish  between 
the $E_{2u}$ and $E_{1u}$ gap functions which was recently much controversial. For $\omega=0.2$ (right column) we may identify certain characteristic wave vectors $\bq_i,\bq_i'$, which are directly related (Fig.~\ref{fig:Fig3}) to the surface DOS of continuum slab states. They are quite different for the three cases, in particular $E_{1u}$ is singled out due to the linear stripe like QPI structures $\parallel q_x$ which are clearly related to scattering between the off-symmetry nodal planes. 

In summary, we have presented a theoretical analysis of QPI spectrum in the 3D 2-component SC \UPT~based on  finite slab approach. The chiral nature and point nodal structure of the three gap candidates ensure the existence of zero energy flat bands within the gap. They lead to one or two Weyl arcs which may provide important selection criterion from zero bias QPI spectrum. At finite bias the surface continuum states lead to different DOS and QPI spectrum associated with the different nodal structure. Together these QPI features provide a direct fingerprint to identify the correct superconducting gap model for \UPT.

{\it Acknowledgments:}
This work was supported by the DFG SPP 1666 ``Topological insulators" (ER463/9-1).
A.A. acknowledges support through  NRF funded by MSIP of Korea (2015R1C1A1A01052411), and by  Max Planck POSTECH / KOREA Research Initiative (No. 2011-0031558) programs through NRF funded by MSIP of Korea. 



\bibliography{References}

\end{document}